\documentclass[twocolumn]{aastex631}

\received{January 12, 2023}
\revised{February 17, 2023}
\accepted{March 10, 2023}

\shorttitle{Identification of Planet in HD 142666}
\shortauthors{Terry et al.}

\begin{document}

\title{Kinematic Evidence of an Embedded Protoplanet in HD 142666 Identified by Machine Learning}

\author[0000-0002-8590-7271]{J. P. Terry}
\affil{Department of Physics and Astronomy, The University of Georgia, Athens, GA 30602, USA.}
\affil{Center for Simulational Physics, The University of Georgia, Athens, GA 30602, USA.}

\author[0000-0002-8138-0425]{C. Hall}
\affil{Department of Physics and Astronomy, The University of Georgia, Athens, GA 30602, USA.} 
\affil{Center for Simulational Physics, The University of Georgia, Athens, GA 30602, USA.}

\author[0000-0002-9426-3789]{S. Abreau}
\affil{Department of Medicine, Division of Cardiology, University of California, San Francisco, CA, 94143, USA.}
\affil{Cardiovascular Research Institute, San Francisco, CA, 94158, USA.}

\author[0000-0002-6222-8102]{S. Gleyzer}
\affil{Department of Physics and Astronomy, The University of Alabama, Tuscaloosa, AL 35487, USA}



\begin{abstract}

Observations of protoplanetary disks have shown that forming exoplanets leave characteristic imprints on the gas and dust of the disk. In the gas, these forming exoplanets cause deviations from Keplerian motion, which can be detected through molecular line observations. Our previous work has shown that machine learning can correctly determine if a planet is present in these disks. Using our machine learning models, we identify strong, localized non-Keplerian motion within the disk HD 142666. Subsequent hydrodynamics simulations of a system with a 5 Jupiter-mass planet at 75 au recreates the kinematic structure. By currently established standards in the field, we conclude that HD 142666 hosts a planet. This work represents a first step towards using machine learning to identify previously overlooked non-Keplerian features in protoplanetary disks.

\end{abstract}

 \keywords{Hydrodynamics --- Radiative transfer --- Accretion disks --- Methods: numerical --- Catalogs --- Planets and satellites: formation}

\section{Introduction} \label{sec:into}

Protoplanetary accretion disks are the sites of planet formation. The newest generation of telescopes, such as the Atacama Large Millimeter/submillimeter Array (ALMA), have unprecedented capabilities for observing protoplanetary disks. For the first time, we can not only resolve disks themselves, but also quantify the motion of the dust and gas within them. Disks display a striking variety of structures such as rings~\citep{hltau2015, dipierro2018}, likely caused by dust trapping due to  forming planets~\citep{pinilla2012,dipierro2015}, and spirals~\citep{perez_2016}, which may be caused by forming planets~\citep[e.g.][]{dong2015} or another mechanism such as gravitational instability~\citep[e.g.][]{donghall2015,hall2018,meru2017}. This new information has greatly advanced our understanding of the processes underlying the formation and evolution of planetary systems.

\par
Planets and physical processes, such as gravitational instability, influence the motion within the disk. This causes the material to deviate from simple Keplerian motion. Comparing the observed motion against purely Keplerian motion provides information on the bodies and processes present in the disk~\citep{hall2020, paneque2021, longarini2021, pinte2022, bae_2022, terry2022_gi}. Non-Keplerian motion has been used to uncover a variety of structures, including localized perturbations associated with gaps and planets~\citep{teague_2018, pinte_2018, pinte_2019, pinte_2020} as predicted by~\citet{perez_2015}.

\par
Kinematic analysis is limited by our ability to accurately identify non-Keplerian motion. The deviations can be small and frequently occur in noisy images. It is therefore not only difficult and slow to identify them, but there is also the strong possibility of overlooking their occurrence. Any signature that is overlooked is a missed opportunity to detect either a forming planet or some other process, such as the GI-Wiggle indicative of gravitational instability~\citep{hall2020} or the vertical shear instability~\citep{barraza_2021}. 

\par 
Machine learning (ML) provides a useful tool for this task. ML has quickly become ubiquitous in both society and the sciences, everything from self-driving cars~\citep{self_driving_cars} to medicine~\citep{Parmar_2015}. Recent efforts in astronomy have made it clear that machine learning is a powerful method even with simulated training data~\citep{jo2019, moller20, alexander_2020}. Machine learning, and in particular computer vision, excels at the analysis of images~\citep{voulodimos2018}. In some cases, it has even been shown to outperform humans~\citep{Zhou2021}. It is therefore naturally suited for application to the noisy datasets in observational astronomy.

\par 
Using ML models developed in a previous work~\citep{terry_ml_2022}, we identify a strong and localized deviation from Keplerian motion in HD 142666. Using the current widely accepted field standard method~\citep{teague_2018, pinte_2018, pinte_2019}, we perform smoothed particle hydrodynamic (SPH) simulations to recreate the kinematic structure of the disk. The agreement is significant when a 5 M$_{J}$ planet is included at 75 au. We conclude that HD 142666 hosts a planet.

\par 
The paper is arranged as follows: Section~\ref{sec:methods} describes the models and simulations used. Section~\ref{sec:results} shows the results of applying the models and simulating the system. Section~\ref{sec:conclusion} gives our conclusions.

\section{Methods} \label{sec:methods}

\subsection{Machine Learning} \label{ssec:models}

We use the ML models described in~\citet{terry_ml_2022} and describe them here for completeness. We use two different architectures: EfficientNetV2~\citep{effnet_v2} and RegNet~\citep{regnet}. All models were made using PyTorch~\citep{pytorch}, albeit with significant modifications to the default models and hyperparameters. We denote these models as EN47, EN61, EN75, RN47, RN61, and RN75. Table~\ref{tab:metrics} gives performance metrics for the models: model accuracy at 50\% and 95\% decision thresholds and the area under the receiver operating characteristic curve (AUC).

\par
The models were trained using synthetic observations from the \texttt{MCFOST}~\citep{mcfost_1, mcfost_2} radiative transfer code. \texttt{MCFOST} inputs were drawn from 1000 \texttt{PHANTOM}~\citep{phantom18} SPH simulations of systems with and without planets~\citep{terry_ml_2022}. Each \texttt{MCFOST} calculation outputs a position-position-velocity cube from $^{13}$CO transition lines ($J=2\rightarrow1$ and $J=3\rightarrow2$). The cubes were convolved spatially and spectrally and noise was added in order to replicate current observational capabilities. 

\par
The model inputs (i.e. radiative transfer outputs) are images of dimension $C \times H \times W$, where $C$ is the number of input channels, $H$ is the height of the image, and $W$ is the width of the image (here, $H=W=600$ pixels). A typical grayscale or RGB image will have $C$=1 or 3, respectively. We instead input an entire position-position-velocity cube.
\par 
Observations vary significantly in the number of channels that cover the disk, but the typical range is between $\approx$40-100. To address this, we train three different implementations of each model, which gives us a total of six models. The difference between each implementation is the number of input velocity channels ($C$=47, 61, or 75). 

\par 
Each model outputs a two-component vector such that the sum of the components is 1, i.e. it has undergone softmax activation~\citep{softmax}. This can be interpreted as the probability that the given input belongs to a certain class, i.e. planet- vs no-planet class. The models also output images of their internal activation structure, which we consider to be the more important output in this context. While the models were not trained to pinpoint the locations of planets \textemdash a job more suited for semantic image segmentation~\citep{segmentation} \textemdash the activation structure can inform us which regions the model finds important when making its classification decision.~\citet{terry_ml_2022} found that the activations were able to highlight velocity channels with non-Keplerian motion in systems that host planet(s).

\par 
To this end, we apply our previously trained models to ALMA data of the HD 142666 system. We inspect the softmax values and activation structures to gain insight into whether a planet might be in the system and, if so, where its signature is the strongest.

\subsection{Observational Data} \label{ssec:data}

The HD 142666 data was taken from the \href{https://almascience.eso.org/almadata/lp/DSHARP/}{DSHARP catalogue}~\citep{dsharp_1, dsharp_2}. 
Data includes $^{12}$CO line emission ($J=2\rightarrow1$) and 1.25 mm continuum images. The system was imaged with a beam with FWHM of $77 \times 61$ mas ($\approx$$11\times9$ au) with an RMS noise of 1.3 mJy beam$^{-1}$; channels have a 0.35 km s$^{-1}$ resolution~\citep{dsharp_1}. Figure~\ref{fig:multi_overlay} shows selected velocity channels overlaid on the continuum. 

\par
The image was cropped to focus on the disk, and a subset of velocity channels was used. The channels were reshaped to $600 \times 600$ pixels and normalized such that all pixel values were between 0 and 1.

\subsection{Hydrodynamical Simulations} \label{ssec:sims}

We run a suite of SPH simulations using \texttt{PHANTOM}, varying the mass of the embedded planet between 1 and 5 M$_{J}$. For each simulation, we create channel maps using \texttt{MCFOST} in the same way that the original training data was made. The kink is approximately 75 au from the center of the disk, so we place a planet at this distance.

\par
System parameters are taken from ~\citet{system_params, dsharp_1, dsharp_2}. The stellar mass, temperature, and radius are 2.0 M$_{\odot}$, 7500 K, and 2.2 R$_{\odot}$, respectively. The disk has a mass of 0.0533 M$_{\odot}$, an inner radius of 1.3 au, and an outer radius of 150 au. The system is inclined at 62 degrees with a position angle of 162 and an azimuth of 72 degrees. It is located 148 pc from Earth.

\par
The SPH outputs are used to create line emission maps to mimic ALMA capabilities. These calculations are done using the \texttt{MCFOST} radiative transfer code~\citep{mcfost_1, mcfost_2}. Each calculation uses $10^8$ photon packets and includes carbon/silicate dust~\citep{Draine1984} with a dust-to-gas ratio of 1:100. The resulting outputs were convolved spatially and spectrally to match the observed line emission resolution.


\begin{table*}

    \centering
    \begin{tabular}{lcccccc}
	\hline
	Value & EN47 & EN61 & EN75 &  RN47 & RN61 & RN75 \\
	\hline
	Accuracy at 50\% cutoff (\%) & $97 \pm 0.5$ & $97 \pm 0.5$ & $93 \pm 0.7$ & $78 \pm 1.1$ & $98 \pm 0.4$ & $95 \pm 0.6$ \\
	Accuracy at 95\% cutoff (\%) & $96 \pm 0.5$ & $94 \pm 0.5$ & $88 \pm 0.9$ & $65 \pm 1.3$ & $96 \pm 0.6$ & $92 \pm 0.7$\\
	AUC & $0.99 \pm 0.002$ & $0.99 \pm 0.003$ & $0.98 \pm 0.003$ & $0.86 \pm 0.010$ & $>0.99 \pm 0.001$ & $0.98 \pm 0.032$\\

    \hline
    \end{tabular}
\caption{Model performance metrics from~\citet{terry_ml_2022}.}
\label{tab:metrics}
\end{table*}

\section{Results and Discussion} \label{sec:results}

\begin{figure*}
    \centering
    \includegraphics[width=\linewidth]{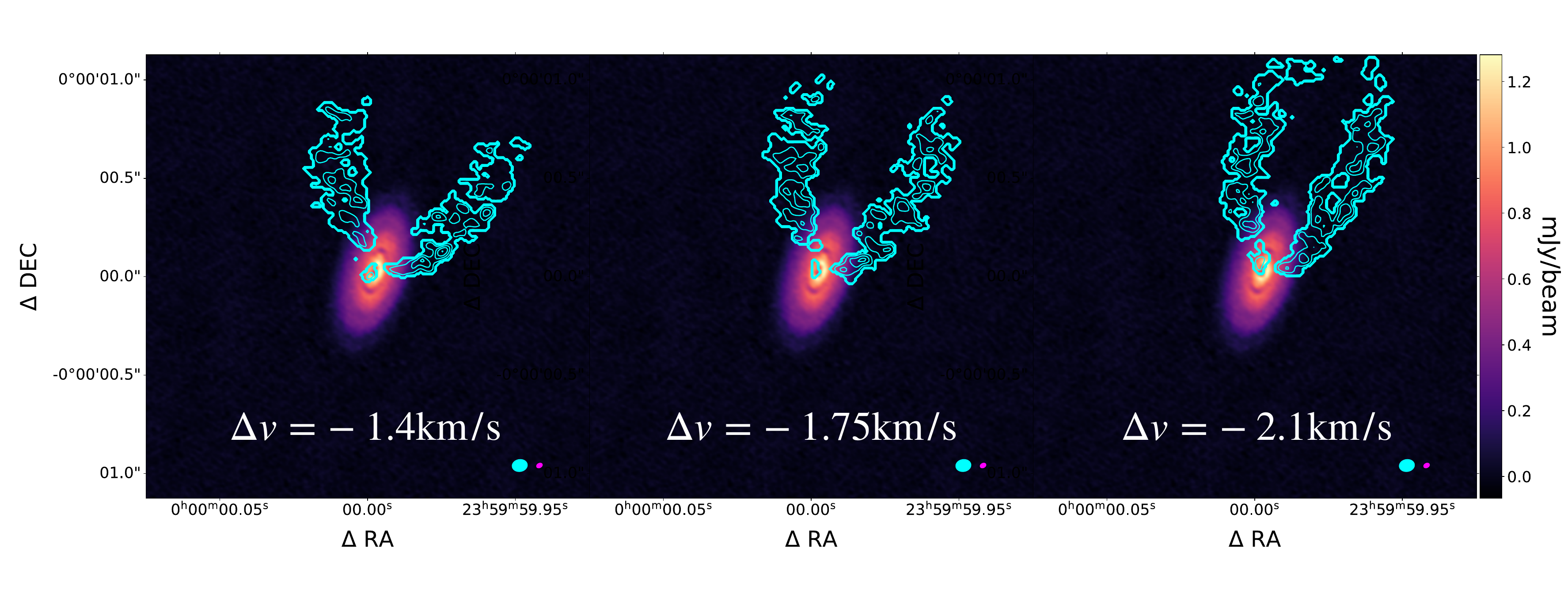}
    \caption{Line emission overlaid on continuum. Left: $\Delta v = -1.4$ km/s channel. Middle: $\Delta v = -1.75$ channel. Right: $\Delta v = -2.1$ channel. The continuum beam is in magenta, and the line emission beam is in cyan.}
    \label{fig:multi_overlay}
\end{figure*}

\begin{figure*}
    \centering
    \includegraphics[width=\linewidth]{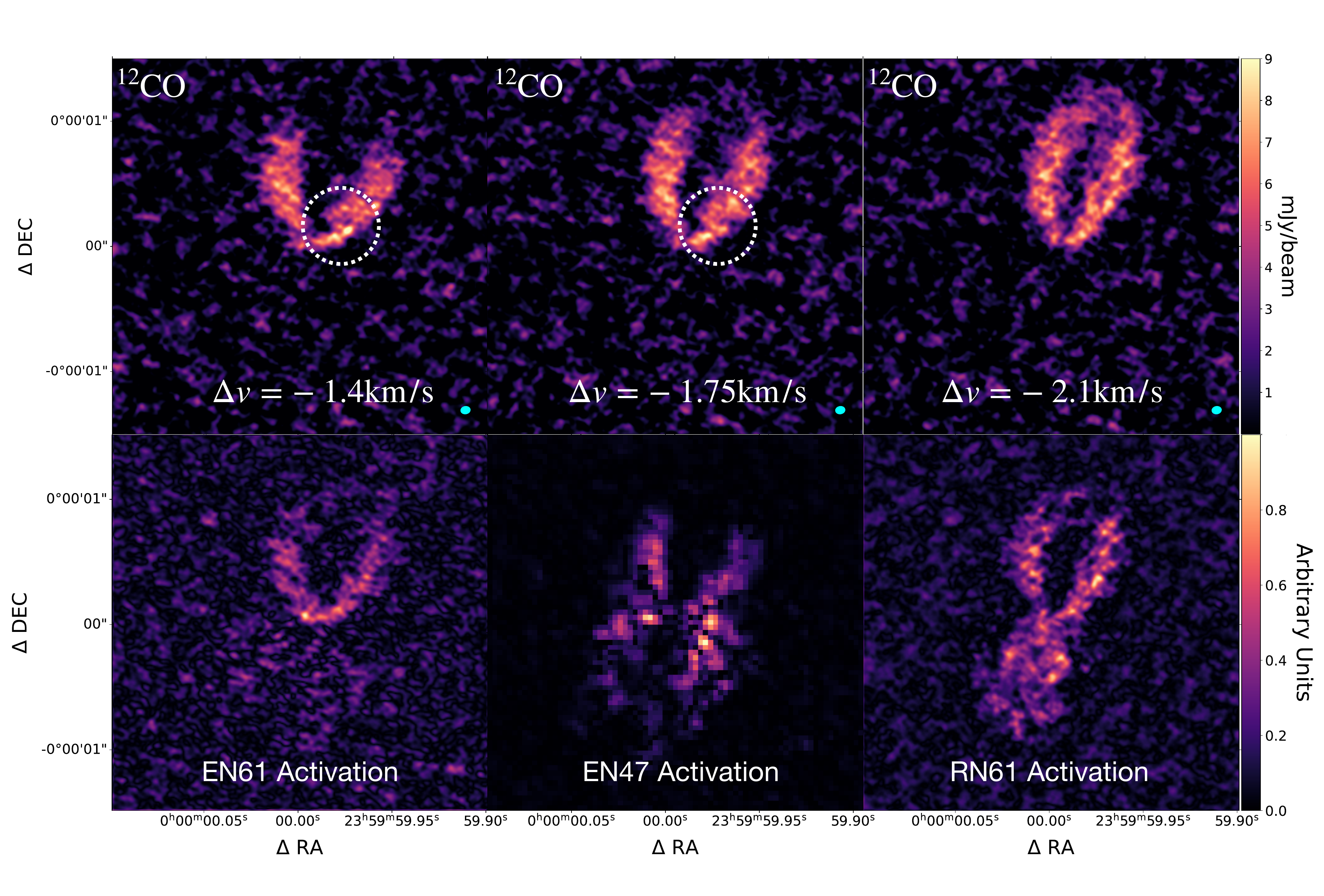}
    \caption{HD 142666 structure ($^{12}$CO: $J=2\rightarrow1$) and activations. Upper left: $\Delta v = -1.4$ km/s channel with kink circled in white. Upper middle: $\Delta v = -1.75$ channel with kink circled in white. Upper right: $\Delta v = -2.1$ channel. Bottom row: selected mean-subtracted activations that roughly correspond to the channels in the upper row. Activations are from three different models (EN61, EN47, and RN61, respectively). Line emission beams are the cyan ellipses in the lower right of the upper row panels.}
    \label{fig:hd_142666_multipanel}
\end{figure*}

\begin{figure*}
    \centering
    \includegraphics[width=\linewidth]{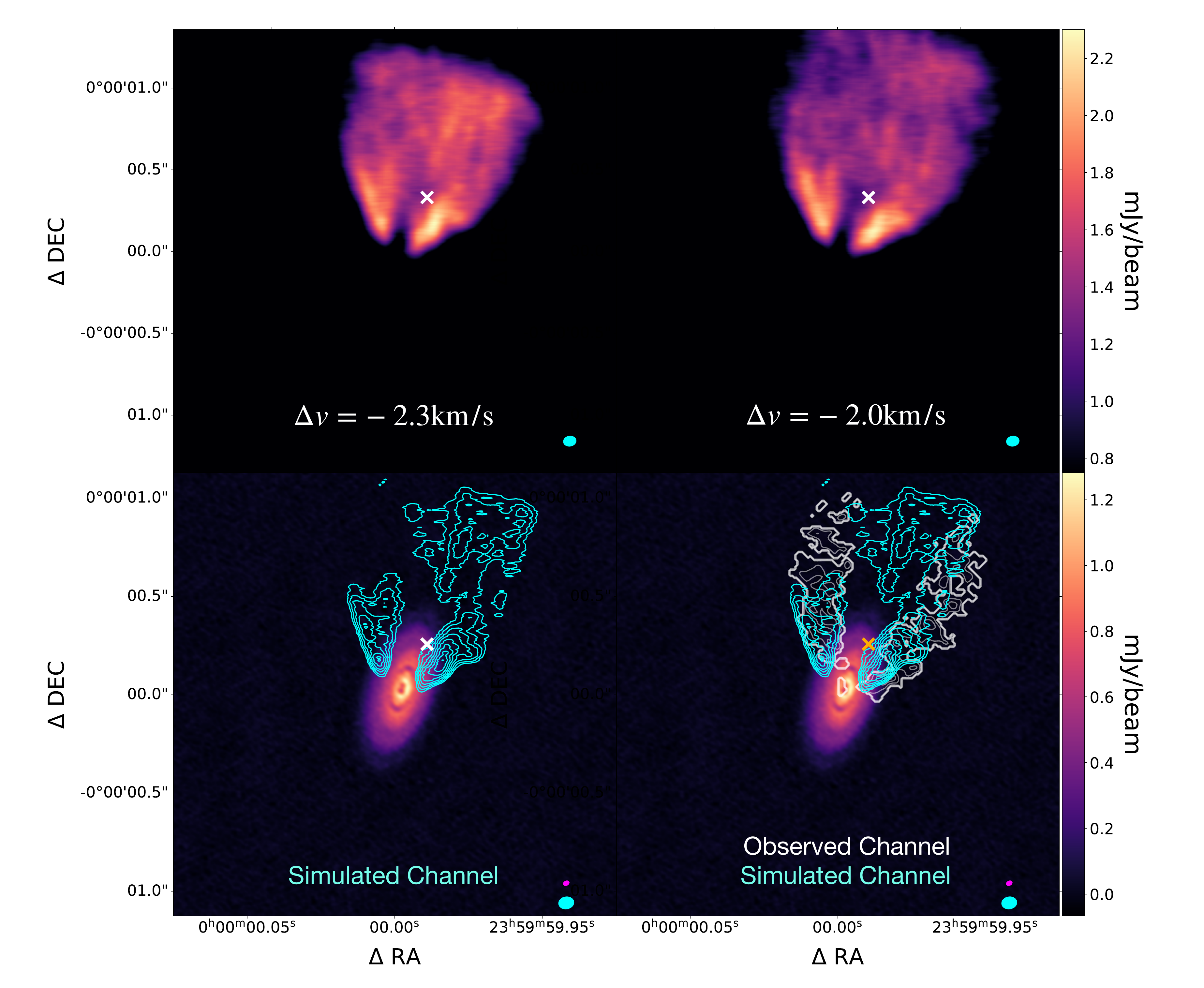}
    \caption{HD 142666 simulation results. Upper left: $\Delta v = -2.3$ km/s channel from the simulation (convolved beam in lower right). Upper right: $\Delta v = -2.0$ km/s channel from the simulation (convolved beam in lower right). Lower left: observed continuum overlaid with contours of simulated $\Delta v = -2.3$ km/s channel. Lower right: observed continuum overlaid with simulated (cyan) and observed (white) channels. Continuum beam is in magenta, and the line emission (simulated and observed) beam is in cyan. The system includes a 5 M$_{J}$ planet at 75 au. The simulated channels have the continuum and background subtracted for clarity. The planet's location is indicated with an x.}
    \label{fig:hd_142666_sim}
\end{figure*}

\par
Figure~\ref{fig:hd_142666_multipanel} shows that HD 142666 has a strong, localized kink that is detected by the ML models. The kink is particularly visible in the upper middle ($\Delta v = -1.75$ km/s) channel. The lower row shows activation structures that roughly correspond to the above channels.

\par
The average softmax value is over 0.84, which means that the models predict the probability that the input for HD 142666 contains a planet to be over 84\%. This prompts further scrutiny of the activations, which we use to determine the most probable channel that contains the kink.  

\par
The strength and localization of the newly identified kink are reminiscent of the kinks in HD 163296 and HD 97048. As with HD 163296, the kink in the gas is outside of the radial extent of the continuum disk. Both of these disks were found to host planets after SPH simulations containing a planet recreated the kinematic structure observed in CO observations~\citep{teague_2018, pinte_2018, pinte_2019}. We apply this same method to HD 142666 to demonstrate that the kink identified by our models is consistent with kinks identified by conventional means in HD 163296 and HD 97048.

\par
 We found that a simulation of a protoplanetary disk with a 5 M$_{J}$ planet most accurately reproduced the observation. Figure~\ref{fig:hd_142666_sim} shows the results. A localized kink in the vicinity of the planet is clear in the upper left panel Figure~\ref{fig:hd_142666_sim} ($\Delta v = -2.3$ km/s). This kink is visible to a lesser extent in the $\Delta v = -2.0$ km/s in the upper right panel 
of Figure~\ref{fig:hd_142666_sim}, which is also the case in Figure~\ref{fig:multi_overlay}. There is strong agreement between this feature and the non-Keplerian channel identified by our models: both display a kink of approximately the same shape and size at approximately the same radial location. This can be seen in the lower left and right panels of Figure~\ref{fig:hd_142666_sim}. Note that the simulation and observation do not display the strongest kink in the same velocity channel. This is simply a relic of the finite temporal resolution of the simulation, which makes it extremely unlikely that the simulation will be saved when the planet is exactly coincident with the observation. The temporal resolution of the simulation was increased to mitigate this effect, but it persists to some extent.

\par
We conclude that HD 142666 hosts a planet. 


\par
We note that our conclusion is confirmed using the same methods described by~\citet{teague_2018, pinte_2018, pinte_2019}. However, what is new about our approach is that the non-Keplerian motion was first identified by ML models, highlighting a protoplanet candidate that had previously been missed upon visual analysis. Verification of the evidence is still done using the same methodology as previous works~\citep{pinte_2018, pinte_2019}. We strongly advocate that this should always be done for any potential discovery.




\subsection{Future Work and Limitations} \label{ssec:limits}

This work shows that machine learning can effectively identify non-Keplerian motion even if it is missed by humans. However, our work can be improved upon. The primary limitation is the fact that localising non-Keplerian motion was not the explicit goal of these models when they were trained. Their purpose was classification without any attempt of segmentation or object detection. Models specifically designed to pinpoint deviations would likely be more effective. Rather than inspecting activation structures \textemdash of which there can be hundreds \textemdash the model would directly output a prediction of the location. This would be a more precise and straightforward method to detect the non-Keplerian signature, but it would not remove the need to perform follow-up simulations. We intend to explore this possibility in future works.

\par
Networks such as PGNets~\citep{pgnets} and DPNNet-2.0~\citep{dpnnet} offer a potentially fruitful route that would increase the accuracy and speed of the analysis of disks and channels highlighted by our models. These networks are designed to infer planetary mass from continuum images. One could use our models to determine if it is likely that a disk hosts a planet and, if so, feed the corresponding continuum images into the secondary networks. The predicted planet mass could then be used as a starting point for followup simulations rather than simply starting an uninformed parameter sweep. This would speed up the verification step of the procedure. Such a pipeline could be useful to explore in future works.

\section{Conclusion} \label{sec:conclusion}

We have applied ML models created by~\citet{terry_ml_2022} to the DSHARP data of HD 142666. All models strongly predict the presence of at least one planet. The activation structures highlight a strong, unreported, and localized kink. An SPH simulation with a 5 M$_{J}$ planet at 75 au is able to recreate the newly identified kinematic structure. By the previously established benchmarks and methods for kinematic planet detection, we conclude that HD 142666 hosts a planet.

\par 
This work demonstrates the utility of applying machine learning to the analysis of protoplanetary disks. By highlighting non-Keplerian features in the disk, ML models are able to guide planet-detection efforts. The signatures of the planet were previously overlooked by human analysts, and the traditional analysis was only performed because of the information given by the models. We anticipate that this method can identify new non-Keplerian features in both existing and future protoplanetary observations.

\section{Acknowledgements}

This paper makes use of the following ALMA data: ADS/JAO.ALMA \#2016.1.00484.L. ALMA is a partnership of ESO (representing its member states), NSF (USA) and NINS (Japan), together with NRC (Canada), MOST and ASIAA (Taiwan), and KASI (Republic of Korea), in cooperation with the Republic of Chile. The Joint ALMA Observatory is operated by ESO, AUI/NRAO and NAOJ. J.T. was a participant in the 2022 Machine Learning for Science (ML4SCI) Google Summer of Code program. S.G. was supported in part by
the National Science Foundation Award No. 2108645. This study was supported in part by resources and technical expertise from the Georgia Advanced Computing Resource Center, a partnership between the University of Georgia’s Office of the Vice President for Research and Office of the Vice President for Information Technology. 



\bibliography{paper}{}

\begin{thebibliography}{}
\expandafter\ifx\csname natexlab\endcsname\relax\def\natexlab#1{#1}\fi
\providecommand{\url}[1]{\href{#1}{#1}}
\providecommand{\dodoi}[1]{doi:~\href{http://doi.org/#1}{\nolinkurl{#1}}}
\providecommand{\doeprint}[1]{\href{http://ascl.net/#1}{\nolinkurl{http://ascl.net/#1}}}
\providecommand{\doarXiv}[1]{\href{https://arxiv.org/abs/#1}{\nolinkurl{https://arxiv.org/abs/#1}}}

\bibitem[{{Alexander} {et~al.}(2020){Alexander}, {Gleyzer}, {McDonough},
  {Toomey}, \& {Usai}}]{alexander_2020}
{Alexander}, S., {Gleyzer}, S., {McDonough}, E., {Toomey}, M.~W., \& {Usai}, E.
  2020, \apj, 893, 15, \dodoi{10.3847/1538-4357/ab7925}

\bibitem[{{ALMA Partnership} {et~al.}(2015){ALMA Partnership}, {Brogan},
  {P{\'e}rez}, {Hunter}, {Dent}, {Hales}, {Hills}, {Corder}, {Fomalont},
  {Vlahakis}, {Asaki}, {Barkats}, {Hirota}, {Hodge}, {Impellizzeri}, {Kneissl},
  {Liuzzo}, {Lucas}, {Marcelino}, {Matsushita}, {Nakanishi}, {Phillips},
  {Richards}, {Toledo}, {Aladro}, {Broguiere}, {Cortes}, {Cortes}, {Espada},
  {Galarza}, {Garcia-Appadoo}, {Guzman-Ramirez}, {Humphreys}, {Jung}, {Kameno},
  {Laing}, {Leon}, {Marconi}, {Mignano}, {Nikolic}, {Nyman}, {Radiszcz},
  {Remijan}, {Rod{\'o}n}, {Sawada}, {Takahashi}, {Tilanus}, {Vila Vilaro},
  {Watson}, {Wiklind}, {Akiyama}, {Chapillon}, {de Gregorio-Monsalvo}, {Di
  Francesco}, {Gueth}, {Kawamura}, {Lee}, {Nguyen Luong}, {Mangum}, {Pietu},
  {Sanhueza}, {Saigo}, {Takakuwa}, {Ubach}, {van Kempen}, {Wootten},
  {Castro-Carrizo}, {Francke}, {Gallardo}, {Garcia}, {Gonzalez}, {Hill},
  {Kaminski}, {Kurono}, {Liu}, {Lopez}, {Morales}, {Plarre}, {Schieven},
  {Testi}, {Videla}, {Villard}, {Andreani}, {Hibbard}, \&
  {Tatematsu}}]{hltau2015}
{ALMA Partnership}, {Brogan}, C.~L., {P{\'e}rez}, L.~M., {et~al.} 2015, \apjl,
  808, L3, \dodoi{10.1088/2041-8205/808/1/L3}

\bibitem[{{Andrews} {et~al.}(2018){Andrews}, {Huang}, {P{\'e}rez}, {Isella},
  {Dullemond}, {Kurtovic}, {Guzm{\'a}n}, {Carpenter}, {Wilner}, {Zhang}, {Zhu},
  {Birnstiel}, {Bai}, {Benisty}, {Hughes}, {{\"O}berg}, \& {Ricci}}]{dsharp_1}
{Andrews}, S.~M., {Huang}, J., {P{\'e}rez}, L.~M., {et~al.} 2018, \apjl, 869,
  L41, \dodoi{10.3847/2041-8213/aaf741}

\bibitem[{{Auddy} {et~al.}(2021){Auddy}, {Dey}, {Lin}, \& {Hall}}]{dpnnet}
{Auddy}, S., {Dey}, R., {Lin}, M.-K., \& {Hall}, C. 2021, \apj, 920, 3,
  \dodoi{10.3847/1538-4357/ac1518}

\bibitem[{{Bae} {et~al.}(2022){Bae}, {Teague}, {Andrews}, {Benisty},
  {Facchini}, {Galloway-Sprietsma}, {Loomis}, {Aikawa}, {Alarc{\'o}n},
  {Bergin}, {Bergner}, {Booth}, {Cataldi}, {Cleeves}, {Czekala}, {Guzm{\'a}n},
  {Huang}, {Ilee}, {Kurtovic}, {Law}, {Gal}, {Liu}, {Long}, {M{\'e}nard},
  {{\"O}berg}, {P{\'e}rez}, {Qi}, {Schwarz}, {Sierra}, {Walsh}, {Wilner}, \&
  {Zhang}}]{bae_2022}
{Bae}, J., {Teague}, R., {Andrews}, S.~M., {et~al.} 2022, \apjl, 934, L20,
  \dodoi{10.3847/2041-8213/ac7fa3}

\bibitem[{{Barraza-Alfaro} {et~al.}(2021){Barraza-Alfaro}, {Flock}, {Marino},
  \& {P{\'e}rez}}]{barraza_2021}
{Barraza-Alfaro}, M., {Flock}, M., {Marino}, S., \& {P{\'e}rez}, S. 2021, \aap,
  653, A113, \dodoi{10.1051/0004-6361/202140535}

\bibitem[{{Bojarski} {et~al.}(2016){Bojarski}, {Del Testa}, {Dworakowski},
  {Firner}, {Flepp}, {Goyal}, {Jackel}, {Monfort}, {Muller}, {Zhang}, {Zhang},
  {Zhao}, \& {Zieba}}]{self_driving_cars}
{Bojarski}, M., {Del Testa}, D., {Dworakowski}, D., {et~al.} 2016, arXiv
  e-prints, arXiv:1604.07316.
\newblock \doarXiv{1604.07316}

\bibitem[{{Dipierro} {et~al.}(2015){Dipierro}, {Price}, {Laibe}, {Hirsh},
  {Cerioli}, \& {Lodato}}]{dipierro2015}
{Dipierro}, G., {Price}, D., {Laibe}, G., {et~al.} 2015, \mnras, 453, L73,
  \dodoi{10.1093/mnrasl/slv105}

\bibitem[{{Dipierro} {et~al.}(2018){Dipierro}, {Ricci}, {P{\'e}rez}, {Lodato},
  {Alexander}, {Laibe}, {Andrews}, {Carpenter}, {Chandler}, {Greaves}, {Hall},
  {Henning}, {Kwon}, {Linz}, {Mundy}, {Sargent}, {Tazzari}, {Testi}, \&
  {Wilner}}]{dipierro2018}
{Dipierro}, G., {Ricci}, L., {P{\'e}rez}, L., {et~al.} 2018, \mnras, 475, 5296,
  \dodoi{10.1093/mnras/sty181}

\bibitem[{{Dong} {et~al.}(2015{\natexlab{a}}){Dong}, {Hall}, {Rice}, \&
  {Chiang}}]{donghall2015}
{Dong}, R., {Hall}, C., {Rice}, K., \& {Chiang}, E. 2015{\natexlab{a}}, \apjl,
  812, L32, \dodoi{10.1088/2041-8205/812/2/L32}

\bibitem[{{Dong} {et~al.}(2015{\natexlab{b}}){Dong}, {Zhu}, {Rafikov}, \&
  {Stone}}]{dong2015}
{Dong}, R., {Zhu}, Z., {Rafikov}, R.~R., \& {Stone}, J.~M. 2015{\natexlab{b}},
  \apjl, 809, L5, \dodoi{10.1088/2041-8205/809/1/L5}

\bibitem[{{Draine} \& {Lee}(1984)}]{Draine1984}
{Draine}, B.~T., \& {Lee}, H.~M. 1984, \apj, 285, 89, \dodoi{10.1086/162480}

\bibitem[{Goodfellow {et~al.}(2016)Goodfellow, Bengio, \& Courville}]{softmax}
Goodfellow, I., Bengio, Y., \& Courville, A. 2016, Deep Learning (MIT Press)

\bibitem[{{Hall} {et~al.}(2018){Hall}, {Rice}, {Dipierro}, {Forgan}, {Harries},
  \& {Alexander}}]{hall2018}
{Hall}, C., {Rice}, K., {Dipierro}, G., {et~al.} 2018, \mnras, 477, 1004,
  \dodoi{10.1093/mnras/sty550}

\bibitem[{{Hall} {et~al.}(2020){Hall}, {Dong}, {Teague}, {Terry}, {Pinte},
  {Paneque-Carre{\~n}o}, {Veronesi}, {Alexander}, \& {Lodato}}]{hall2020}
{Hall}, C., {Dong}, R., {Teague}, R., {et~al.} 2020, \apj, 904, 148,
  \dodoi{10.3847/1538-4357/abac17}

\bibitem[{{Huang} {et~al.}(2018){Huang}, {Andrews}, {Dullemond}, {Isella},
  {P{\'e}rez}, {Guzm{\'a}n}, {{\"O}berg}, {Zhu}, {Zhang}, {Bai}, {Benisty},
  {Birnstiel}, {Carpenter}, {Hughes}, {Ricci}, {Weaver}, \&
  {Wilner}}]{dsharp_2}
{Huang}, J., {Andrews}, S.~M., {Dullemond}, C.~P., {et~al.} 2018, \apjl, 869,
  L42, \dodoi{10.3847/2041-8213/aaf740}

\bibitem[{{Jo} \& {Kim}(2019)}]{jo2019}
{Jo}, Y., \& {Kim}, J.-h. 2019, \mnras, 489, 3565,
  \dodoi{10.1093/mnras/stz2304}

\bibitem[{{Longarini} {et~al.}(2021){Longarini}, {Lodato}, {Toci}, {Veronesi},
  {Hall}, {Dong}, \& {Patrick Terry}}]{longarini2021}
{Longarini}, C., {Lodato}, G., {Toci}, C., {et~al.} 2021, \apjl, 920, L41,
  \dodoi{10.3847/2041-8213/ac2df6}

\bibitem[{{Meru} {et~al.}(2017){Meru}, {Juh{\'a}sz}, {Ilee}, {Clarke},
  {Rosotti}, \& {Booth}}]{meru2017}
{Meru}, F., {Juh{\'a}sz}, A., {Ilee}, J.~D., {et~al.} 2017, \apjl, 839, L24,
  \dodoi{10.3847/2041-8213/aa6837}

\bibitem[{Minaee {et~al.}(2022)Minaee, Boykov, Porikli, Plaza, Kehtarnavaz, \&
  Terzopoulos}]{segmentation}
Minaee, S., Boykov, Y., Porikli, F., {et~al.} 2022, IEEE Transactions on
  Pattern Analysis and Machine Intelligence, 44, 3523,
  \dodoi{10.1109/TPAMI.2021.3059968}

\bibitem[{{M{\"o}ller} \& {de Boissi{\`e}re}(2020)}]{moller20}
{M{\"o}ller}, A., \& {de Boissi{\`e}re}, T. 2020, \mnras, 491, 4277,
  \dodoi{10.1093/mnras/stz3312}

\bibitem[{{Paneque-Carre{\~n}o} {et~al.}(2021){Paneque-Carre{\~n}o},
  {P{\'e}rez}, {Benisty}, {Hall}, {Veronesi}, {Lodato}, {Sierra}, {Carpenter},
  {Andrews}, {Bae}, {Henning}, {Kwon}, {Linz}, {Loinard}, {Pinte}, {Ricci},
  {Tazzari}, {Testi}, \& {Wilner}}]{paneque2021}
{Paneque-Carre{\~n}o}, T., {P{\'e}rez}, L.~M., {Benisty}, M., {et~al.} 2021,
  \apj, 914, 88, \dodoi{10.3847/1538-4357/abf243}

\bibitem[{{Parmar} {et~al.}(2015){Parmar}, {Grossmann}, {Bussink}, {Lambin}, \&
  {Aerts}}]{Parmar_2015}
{Parmar}, C., {Grossmann}, P., {Bussink}, J., {Lambin}, P., \& {Aerts}, H.
  J.~W.~L. 2015, Scientific Reports, 5, 13087, \dodoi{10.1038/srep13087}

\bibitem[{Paszke {et~al.}(2019)Paszke, Gross, Massa, Lerer, Bradbury, Chanan,
  Killeen, Lin, Gimelshein, Antiga, Desmaison, Kopf, Yang, DeVito, Raison,
  Tejani, Chilamkurthy, Steiner, Fang, Bai, \& Chintala}]{pytorch}
Paszke, A., Gross, S., Massa, F., {et~al.} 2019, in Advances in Neural
  Information Processing Systems, ed. H.~Wallach, H.~Larochelle,
  A.~Beygelzimer, F.~d\textquotesingle Alch\'{e}-Buc, E.~Fox, \& R.~Garnett,
  Vol.~32 (Curran Associates, Inc.), \dodoi{10.48550/arXiv.1912.01703}

\bibitem[{{P{\'e}rez} {et~al.}(2016){P{\'e}rez}, {Carpenter}, {Andrews},
  {Ricci}, {Isella}, {Linz}, {Sargent}, {Wilner}, {Henning}, {Deller},
  {Chandler}, {Dullemond}, {Lazio}, {Menten}, {Corder}, {Storm}, {Testi},
  {Tazzari}, {Kwon}, {Calvet}, {Greaves}, {Harris}, \& {Mundy}}]{perez_2016}
{P{\'e}rez}, L.~M., {Carpenter}, J.~M., {Andrews}, S.~M., {et~al.} 2016,
  Science, 353, 1519, \dodoi{10.1126/science.aaf8296}

\bibitem[{{Perez} {et~al.}(2015){Perez}, {Dunhill}, {Casassus}, {Roman},
  {Szul{\'a}gyi}, {Flores}, {Marino}, \& {Montesinos}}]{perez_2015}
{Perez}, S., {Dunhill}, A., {Casassus}, S., {et~al.} 2015, \apjl, 811, L5,
  \dodoi{10.1088/2041-8205/811/1/L5}

\bibitem[{{Pinilla} {et~al.}(2012){Pinilla}, {Birnstiel}, {Ricci}, {Dullemond},
  {Uribe}, {Testi}, \& {Natta}}]{pinilla2012}
{Pinilla}, P., {Birnstiel}, T., {Ricci}, L., {et~al.} 2012, \aap, 538, A114,
  \dodoi{10.1051/0004-6361/201118204}

\bibitem[{{Pinte} {et~al.}(2009){Pinte}, {Harries}, {Min}, {Watson},
  {Dullemond}, {Woitke}, {M{\'e}nard}, \& {Dur{\'a}n-Rojas}}]{mcfost_2}
{Pinte}, C., {Harries}, T.~J., {Min}, M., {et~al.} 2009, \aap, 498, 967,
  \dodoi{10.1051/0004-6361/200811555}

\bibitem[{{Pinte} {et~al.}(2006){Pinte}, {M{\'e}nard}, {Duch{\^e}ne}, \&
  {Bastien}}]{mcfost_1}
{Pinte}, C., {M{\'e}nard}, F., {Duch{\^e}ne}, G., \& {Bastien}, P. 2006, \aap,
  459, 797, \dodoi{10.1051/0004-6361:20053275}

\bibitem[{{Pinte} {et~al.}(2022){Pinte}, {Teague}, {Flaherty}, {Hall},
  {Facchini}, \& {Casassus}}]{pinte2022}
{Pinte}, C., {Teague}, R., {Flaherty}, K., {et~al.} 2022, arXiv e-prints,
  arXiv:2203.09528.
\newblock \doarXiv{2203.09528}

\bibitem[{{Pinte} {et~al.}(2018){Pinte}, {Price}, {M{\'e}nard}, {Duch{\^e}ne},
  {Dent}, {Hill}, {de Gregorio-Monsalvo}, {Hales}, \& {Mentiplay}}]{pinte_2018}
{Pinte}, C., {Price}, D.~J., {M{\'e}nard}, F., {et~al.} 2018, \apjl, 860, L13,
  \dodoi{10.3847/2041-8213/aac6dc}

\bibitem[{{Pinte} {et~al.}(2019){Pinte}, {van der Plas}, {M{\'e}nard}, {Price},
  {Christiaens}, {Hill}, {Mentiplay}, {Ginski}, {Choquet}, {Boehler},
  {Duch{\^e}ne}, {Perez}, \& {Casassus}}]{pinte_2019}
{Pinte}, C., {van der Plas}, G., {M{\'e}nard}, F., {et~al.} 2019, Nature
  Astronomy, 3, 1109, \dodoi{10.1038/s41550-019-0852-6}

\bibitem[{{Pinte} {et~al.}(2020){Pinte}, {Price}, {M{\'e}nard}, {Duch{\^e}ne},
  {Christiaens}, {Andrews}, {Huang}, {Hill}, {van der Plas}, {Perez}, {Isella},
  {Boehler}, {Dent}, {Mentiplay}, \& {Loomis}}]{pinte_2020}
{Pinte}, C., {Price}, D.~J., {M{\'e}nard}, F., {et~al.} 2020, \apjl, 890, L9,
  \dodoi{10.3847/2041-8213/ab6dda}

\bibitem[{{Price} {et~al.}(2018){Price}, {Wurster}, {Tricco}, {Nixon},
  {Toupin}, {Pettitt}, {Chan}, {Mentiplay}, {Laibe}, {Glover}, {Dobbs},
  {Nealon}, {Liptai}, {Worpel}, {Bonnerot}, {Dipierro}, {Ballabio}, {Ragusa},
  {Federrath}, {Iaconi}, {Reichardt}, {Forgan}, {Hutchison}, {Constantino},
  {Ayliffe}, {Hirsh}, \& {Lodato}}]{phantom18}
{Price}, D.~J., {Wurster}, J., {Tricco}, T.~S., {et~al.} 2018, \pasa, 35, e031,
  \dodoi{10.1017/pasa.2018.25}

\bibitem[{{Rubinstein} {et~al.}(2018){Rubinstein}, {Mac{\'\i}as}, {Espaillat},
  {Zhang}, {Calvet}, \& {Robinson}}]{system_params}
{Rubinstein}, A.~E., {Mac{\'\i}as}, E., {Espaillat}, C.~C., {et~al.} 2018,
  \apj, 860, 7, \dodoi{10.3847/1538-4357/aabfba}

\bibitem[{Tan \& Le(2021)}]{effnet_v2}
Tan, M., \& Le, Q. 2021, in Proceedings of Machine Learning Research, Vol. 139,
  Proceedings of the 38th International Conference on Machine Learning, ed.
  M.~Meila \& T.~Zhang (PMLR), 10096--10106, \dodoi{10.48550/arXiv.2104.00298}

\bibitem[{{Teague} {et~al.}(2018){Teague}, {Bae}, {Bergin}, {Birnstiel}, \&
  {Foreman-Mackey}}]{teague_2018}
{Teague}, R., {Bae}, J., {Bergin}, E.~A., {Birnstiel}, T., \& {Foreman-Mackey},
  D. 2018, \apjl, 860, L12, \dodoi{10.3847/2041-8213/aac6d7}

\bibitem[{{Terry} {et~al.}(2022{\natexlab{a}}){Terry}, {Hall}, {Abreau}, \&
  {Gleyzer}}]{terry_ml_2022}
{Terry}, J.~P., {Hall}, C., {Abreau}, S., \& {Gleyzer}, S. 2022{\natexlab{a}},
  \apj, 941, 192, \dodoi{10.3847/1538-4357/aca477}

\bibitem[{{Terry} {et~al.}(2022{\natexlab{b}}){Terry}, {Hall}, {Longarini},
  {Lodato}, {Toci}, {Veronesi}, {Paneque-Carre{\~n}o}, \&
  {Pinte}}]{terry2022_gi}
{Terry}, J.~P., {Hall}, C., {Longarini}, C., {et~al.} 2022{\natexlab{b}},
  \mnras, 510, 1671, \dodoi{10.1093/mnras/stab3513}

\bibitem[{Voulodimos {et~al.}(2018)Voulodimos, Doulamis, Doulamis,
  Protopapadakis, \& Andina}]{voulodimos2018}
Voulodimos, A., Doulamis, N., Doulamis, A., Protopapadakis, E., \& Andina, D.
  2018, Neuroscience, 7068349, \dodoi{10.1155/2018/7068349}

\bibitem[{Xu {et~al.}(2022)Xu, Pan, Pan, Hoi, Yi, \& Xu}]{regnet}
Xu, J., Pan, Y., Pan, X., {et~al.} 2022, IEEE Transactions on Neural Networks
  and Learning Systems, 1, \dodoi{10.1109/TNNLS.2022.3158966}

\bibitem[{{Zhang} {et~al.}(2022){Zhang}, {Zhu}, \& {Kang}}]{pgnets}
{Zhang}, S., {Zhu}, Z., \& {Kang}, M. 2022, \mnras, 510, 4473,
  \dodoi{10.1093/mnras/stab3502}

\bibitem[{{Zhou} {et~al.}(2021){Zhou}, {Yang}, {Yu}, {Liu}, {Duan}, {Weng},
  {Chen}, {Liang}, {Fang}, {Zhou}, {Ju}, {Luo}, {Guo}, {Ma}, {Xie}, {Wang}, \&
  {Zhou}}]{Zhou2021}
{Zhou}, W., {Yang}, Y., {Yu}, C., {et~al.} 2021, Nature Communications, 12,
  1259, \dodoi{10.1038/s41467-021-21466-z}

\end{thebibliography}
\bibliographystyle{aasjournal}



\end{document}